\documentclass[preprint,aps,eqsecnum,nofootinbib]{revtex4}
\usepackage{epsfig}

\newcommand{\be}{\begin{equation}}
\newcommand{\ee}{\end{equation}}
\newcommand{\bea}{\begin{eqnarray}}
\newcommand{\beas}{\begin{eqnarray*}}
\newcommand{\eea}{\end{eqnarray}}
\newcommand{\eeas}{\end{eqnarray*}}
\newcommand{\ba}{\begin{array}}
\newcommand{\ea}{\end{array}}
\def\ls{\mathrel{\lower4pt\vbox{\lineskip=0pt\baselineskip=0pt
           \hbox{$<$}\hbox{$\sim$}}}}
\def\gs{\mathrel{\lower4pt\vbox{\lineskip=0pt\baselineskip=0pt
           \hbox{$>$}\hbox{$\sim$}}}}

\begin{document}



\title{Radion Cosmology in Theories with Universal Extra Dimensions}
\author{Anupam Mazumdar~$^{1}$,~R. N. Mohapatra~$^{2}$, and
A. P\'erez-Lorenzana~$^{3}$}
\affiliation{$^{1}$~CHEP, McGill University, 3600 University Road, Montr\'eal,
Qu\'ebec, H3A 2T8, Canada\\
$^{2}$~Department of Physics, University of Maryland,
College Park, MD 20742, USA\\
$^{3}$~Departamento de F\'{\i}sica,
Centro de Investigaci\'on y de Estudios Avanzados del I.P.N.\\
Apdo. Post. 14-740, 07000, M\'exico, D.F., M\'exico}

\date{October, 2001}

\begin{abstract}
We discuss cosmology of models with universal extra dimensions, 
where the Standard Model degrees of freedom live in a $4+n$ dimensional 
brane, with $n$ compact and small extra spatial dimensions. In these
models, the simplest
way to obtain the conventional 4-dimensional Planck scale starting with a 
 low string scale is to have also some larger extra dimensions, where only
gravity propagates. In such theories, dimensional reduction generically 
leads to at least two radion fields, one associated
with the total volume of the extra spatial dimensions, and the other with
the ratio 
of the sizes of small and large extra dimensions. In this paper,
we discuss the impact of the radion fields on cosmology. We emphasize various
aspects of radion physics such as radion coupling to the Standard Model
fields, bare 
and dressed radion masses during inflation, dynamical stabilization of 
radions during and after inflation, radion decay life time and its late 
dominance in thermal history of the Universe as well as its quantum
fluctuations during inflation. We argue that models where the radion plays
the role of an inflaton or the inflaton is a brane scalar field, run into
problems. We then present a successful inflation model with bulk scalar
fields that seems to have all the desired properties. We also briefly
discuss the possibility of radion as a cold dark matter candidate.
\end{abstract}

\maketitle
\section{Introduction}

The idea that the universe may have extra spatial dimensions
has been under consideration for many years. The earliest thoughts date 
back to the  works of Kaluza and Klein in the 1920's. Such a 
possibility has been lately reinforced by the advent of string theory 
which is believed to be the most attractive candidate for a theory
of gravity and gauge interactions~\cite{strings}, where the simplest  
realization requires at least $10$  dimensions to be consistent. 
The extra dimensions are usually assumed to be compact with the sizes 
of the order of the fundamental  Planck scale, 
$l_{p}= M_{p}^{-1}=(2.436\times 10^{18}~({\rm GeV})^{-1}\sim 10^{-33}$~cm. 
However recent developments~\cite{witten}
in this field have shown the prospects of having large compact 
dimensions~\cite{early1}. The largest possible compact dimension
could be as large as a few micrometers~\cite{nima0}. In the simplest
incarnation it is believed that the Standard Model (SM) fields live in a 
four dimensional hypersurface or brane embedded in a higher 
dimensional space, which is known as the bulk, while gravity can propagate 
in all dimensions (for some earlier ideas see also Refs.~\cite{early2}).

In these models gravitational interactions appear to be stronger 
when probing distances below the size of the extra dimensions.
The actual fundamental scale of gravity, $M_*$, is no longer 
the (now effective) 4 dimensional  Planck scale, 
but it is rather related to this by~\cite{nima0}
\be
M_{p}= M_{\ast}^{2+d}\cdot {\rm vol}\,;
\label{nima0}
\ee
where ${\rm vol}$ denotes the volume of the $d$ compact 
spatial dimensions. Current measurements on the precision 
of the Newton's law at small distances impose an upper bound on 
the size of the extra dimensions at a level of $100$ 
microns~\cite{expgrav}. For $d=2$ it has been pointed out 
that the fundamental scale of quantum gravity could be as 
low as few ${\cal O}(\rm TeV)$, which will have interesting implications
for
collider physics~\cite{lykken,colliders,cdf}. From this perspective 
these theories have an edge of being testable in future terrestrial and 
extra terrestrial experiments. The models with extra dimensions also 
provide an unique  geometric insight to the naturalness problem in physics,
which is one of the greatest reasons for studying various aspects of these
theories in particle physics and in cosmology and astrophysics. Actually the
stringent bounds on the size of the extra dimensions, or equivalently on the 
fundamental scale have been obtained from the astrophysical implications
of the excited states of graviton, $M_*> 1000$~TeV~\cite{hannestad}.
However very recent study reveals that such a bound can be weakened 
down to few TeV when the momenta associated with the extra dimensions are
broken~\cite{nussinov}.

Another class of extra dimension models proposes that the
SM degrees of freedom also propagate in some of the extra 
dimensions; examples of such studies can be found in 
Refs.~\cite{powerlaw,running,appelquist,ued1,borghini,ued,nup,ued2,ued3}.
In this scenario, each SM particle is accompanied by a 
Kaluza Klein (KK) tower, with masses spaced by the inverse 
size of the extra dimensions in which they reside.  Since there is no 
evidence of a KK tower associated with any SM particle yet, 
the size of such extra dimensions has to be smaller compared to the
inverse of the 
electroweak scale, $\sim {\rm TeV}^{-1}$. These models go by the
name of universal extra dimension models (UED). Amongst the
interesting 
features of these models is the possibility that they can be tested 
in near future in collider 
experiments through direct production of the excited KK 
states of the standard model particles~\cite{appelquist}. Furthermore in
six dimensions, such  scenarios acquire the additional virtue that they 
can provide a natural explanation for the 
number of generations~\cite{ued1,borghini}, and  an understanding of  
proton  stability~\cite{ued,nup} along with the lightness of the 
observed neutrino masses~\cite{nup,ued2}, despite the absence of any large
mass scales in the theory. These models are therefore phenomenologically
more sound than models where the standard model resides in 3+1 dimensional
Minkowski brane.

Cosmology of models with large extra dimensions, where 
only gravity propagates in the bulk has received a great deal of 
attention, from building inflationary models 
\cite{Manyinf,Mohapatra,Lorenzana1,Green,matsuda,king} to studying moduli
problem\cite{Csaki}, moduli stabilization~\cite{Lorenzana1},
thermal history~\cite{Benakli}, and baryogenesis~\cite{Lorenzana2}.
In this paper our main aim is to address similar cosmological 
issues in the class of models where both standard model particles and
gravity propagate in the bulk (or the UED models).

For simplicity we consider the $n$ extra spatial dimensions where the
standard model particles propagate,
compactified on orbifold-ed circles with a common size $r$. 
The power law running of the gauge couplings in these 
models~\cite{powerlaw,running} and the need for keeping the 
theory from violating unitarity bounds~\cite{appelquist}, 
suggest that $r$ should not be much larger than about 10-100 times the 
inverse fundamental scale ($M_*^{-1}$), i.e.
$M_{\ast}r\sim {\cal O}(10)-{\cal O}(100)$. Explaining the weakness of
gravity in such theories (where $M_*$ of order of 10 TeV), requires that
there must be large extra dimensions orthogonal to the 4+n
already considered~\cite{nima0} (i.e. orthogonal to the 
SM brane).

In the simplest scenario, we consider UED models comprising two
sets of asymmetric extra spatial dimensions: $n$-small dimensions, 
of a size $r$ just around a factor of ten to hundred times above
the fundamental scale $M_{\ast}^{-1}$,  and $p$-larger spatial 
dimensions of size $R$. Note that the SM fields are glued to 
$4+n$ dimensions only, while gravity is propagating in the entire 
bulk. Therefore the effective four dimensional Planck mass $M_p$ is given
by
 \be
M_{p}^2\equiv M_{\ast}^{2+n+p}~r^n~R^p\,.
 \label{add1}
 \ee
For $M_{\ast}$ as low as few TeV, there should be at least two large extra
spatial dimensions of sub millimeter size in order to yield the correct
four dimensional Planck scale.

In the above mentioned set-up, upon dimensional reduction, there are 
at least two  radions.  One  describes the variations of the whole volume
of the extra dimensions. We will call it the volume radion. The
other,
here called the shape mode,  describes the relative 
expansion and contraction of the two sets of extra dimensions. 
In this paper we will
take a phenomenological approach in determining the masses of the two 
radions, which could be either light or heavy compared to cosmological 
scale, which is usually taken to be the Hubble parameter. We will show 
that these radions play an extremely important role in determining the 
fate of the early Universe in UED models.

The early universe cosmology is incomplete if it cannot reproduce the
successes of the hot big bang model, e.g. synthesis of light elements 
at a scale $\sim 1$~MeV and their present day observed abundance. However 
this cannot be reached if the universe prior to this era were not already 
filled by the SM degrees of freedom, and provided there were no excess 
generation of entropy after big bang nucleosynthesis (BBN), which would 
otherwise dilute the initial abundance. This poses a challenge 
to any particle physics theory with large extra dimensions. In this regard
UED  models are no exception. They come with light scalar fields and
therefore 
it is paramount that we study the dynamics and  decay life time of 
these radions. The late oscillations and eventual decay of the radions 
can dominate thermal history of the universe, which can pose a serious 
threat to BBN. This is analogous to the string moduli problem. 
Therefore it is important that radions either decay before 
BBN, via their couplings to the SM fields, or they are sufficiently light 
to survive the age of the universe. If they survive late then they can 
be regarded as candidates for cold dark matter. 
In an ideal situation it would be  possible to stabilize both the 
volume and the shape modes during inflation and/or right after the 
end of inflation. This is a major issue which we will deal 
in some detail in our paper. We argue that only the volume
radion can be stabilized dynamically during inflation, 
because it always couple  
to the inflaton, whereas  the shape mode remains
uncoupled to it. Besides this, there is another important difference:
the volume radion governs the four dimensional Newton's constant, therefore
it is important that we stabilize this radion before BBN. There are already 
stringent bounds on variations on the Newton's constant during 
BBN~\cite{clifford}. On the other hand the shape mode can remain 
light and dynamically active field. We will find that it must
also be stabilized before BBN. The shape mode can give rise to the 
running of the SM coupling constants. 

Furthermore, we also describe a successful inflationary
scenario in UED model. We will discuss three possibilities, radion 
driven inflation, brane field driven inflation and the bulk field driven 
inflation and assess the relative merits of the different scenarios.

Light scalars in cosmology are sometimes regarded
as boon if they can sustain a flat potential with a large vacuum energy
since they can then give rise to accelerated expanding phase of the
universe, the inflation phase. They can also dilute all energy densities
except the quantum
fluctuations which can be imprinted on the cosmic microwave background 
radiation. The radions in UED models could be responsible for generating
adiabatic density fluctuations seen by COBE~\cite{COBE}, 
BOOMERANG~\cite{BOOMERANG}, MAXIMA~\cite{MAXIMA}, DASI~\cite{pryke01}
and WMAP~\cite{WMAP1,WMAP}. In this paper we propose that even if the radions
are not the ideal candidate for inflaton, they can still be responsible for
reheating the universe with the SM degrees of freedom and also converting
its fluctuations into adiabatic modes.

This paper is organized as follows. We begin by introducing our 
general set-up and performing a dimensional reduction of the 
gravitational part of the action in order to identify the radion fields. 
In the subsequent sections we devote our discussion on general aspects 
of the radion physics; we introduce the radion to matter couplings in 
section 3; we discuss initial conditions of the radion vacua before the 
onset of inflation in section 4. In section 5, we comment on the problems
of dealing 
with the radion (stabilization) potential and the radion mass, which 
seem to have an important effect on the dynamics of inflation and 
thermal history of the universe. Then we discuss the decay rate and 
life time of the radion fields. We begin section 6 with the analysis 
of the potential role of the radions in the early Universe. First 
we discuss the case when the radion acts as an inflaton. Then we 
extend our scenario by including a brane or a bulk inflaton. Then we discuss 
a possible role of radions to generate SM relativistic degrees of freedom 
along with the adiabatic density perturbations. We conclude our discussion
by mentioning the possibility for the radions to become a candidate for
the cold dark matter.


\section{The Set-up, dimensional reduction and the two radions}

Let us consider the following metric in $4+ d$ dimensions
 \be
 ds^2 = G_{AB}dx^A dx^B =  g_{\mu\nu}dx^\mu dx^\nu - h_{ij} dy^i dy^j\,,
 \ee
with $A,B$ running  over $\mu$ and $i$, where  $\mu= 0,1,2,3$ labels
our usual four dimensions and $i,j = 1, \dots, d$ are the total $d$ extra
dimensions. The coordinates $y^i$ take values in the interval $[0,1]$.
Note that since we have taken $y^i$ to be dimensionless, our $h_{ij}$ has
length dimension two and $h^{ij}$ has $L^{-2}$. 

Also note that we are not considering the presence of  vector-like
connection 
$A_\mu^i$ pieces, which are common in Kaluza Klein theories. This is  
because we are only interested in the zero mode part of the metric, after
orbifolding the extra spatial dimensions. We have $A_\mu^i$ odd under
parity
transformation and therefore it vanishes at the zero mode level. By 
reducing the $4+d$ scalar curvature term, we obtain at zero mode level 
the effective action
 \bea
 S&=&
 {1\over \kappa_{\ast}^2}\, \int\!d^4x\, d^dy\, \sqrt{-G}\, R[G] 
 \nonumber  \\[1ex]
 &=&
 {1\over \kappa^2}\, \int\!d^4x\,\sqrt{-g}\, {\sqrt{h}\over vol_0}\, \left \{
    R[g] - {1\over 4} \partial_\mu h^{ij}\, \partial^\mu h_{ij}
 -{1\over 4}h^{ij}\partial_\mu h_{ij}\cdot h^{kl}\partial^\mu h_{kl}
 \right \}\,,
 \label{rg1}
 \eea
where $1/\kappa_{\ast}^2 = M_{\ast}^{2+d}/2$ defines the true 
fundamental gravity scale, while  $1/\kappa^2 = M_{p}^2/2$, 
defines the effective $4$ dimensional Planck scale, which is 
related to the fundamental scale by a volume suppression, 
$\kappa_{\ast}^2 = \kappa^2 \cdot vol_0$, see Eq.~(\ref{add1}). 
Here $vol_0$ is the stabilized volume of the extra space and corresponds
to stabilized values for the $h_{ij}$.

In order to obtain the $4$ dimensional scalar curvature term in 
a canonical  form, we have to perform a conformal transformation 
on the metric,
 \be
 g_{\mu\nu}\rightarrow e^{2\varphi}g_{\mu\nu}\,,
 \label{conformal}
 \ee
designed to cancel the extra $\sqrt{h}/vol_0$ coefficient of $R[g]$ in
Eq.~(\ref{rg1}). We take $\varphi$ such that
 \be
 e^{2\varphi}\, \sqrt{h}/vol_0 = 1\,.
 \label{ephih}
 \ee
The action in Eq.~(\ref{rg1}) is then transformed into
 \be
 S=
 { 1\over \kappa^2}\, \int\!d^4x\,\sqrt{-g}\, \left \{
    R[g] - {1\over 4} \partial_\mu h^{ij}\, \partial^\mu h_{ij}
 + {1\over 8}h^{ij}\partial_\mu h_{ij}\cdot h^{kl}\partial^\mu h_{kl}
 \right \}\,.
 \label{rg2}
 \ee
Next for the four dimensional part of the metric, $g_{\mu\nu}$, we will 
assume the standard  Friedman-Robertson-Walker metric with a flat geometry
i.e.  \be
    g_{\mu\nu} = diag( 1, -a(t),-a(t), -a(t))\,,
 \ee
for an isotropic and homogeneous (brane) universe, 
whereas we  consider a diagonal form for the $h$ part of the metric:
 \be
 h_{ij} \leftrightarrow diag\{
  {\stackrel{n-{\rm times}}{\overbrace{ h_1^2, \dots,h_1^2}}},
  {\stackrel{p-{\rm times}}{\overbrace{ h_2^2, \dots, h_2^2}}}
  \}\,,
\label{hij}
 \ee
where we have assumed $n$ ($p$) of the $d$ extra dimensions to be small
(large) with common physical sizes, $h_1(t)$ and  $h_2(t)$ respectively. 
Obviously $d= n+p$, and the physical volume of the
extra space is given as
 \[
 vol_{\rm phys}=\sqrt{h}= h_1^n h_2^p~.
 \]
At present, the physical sizes of the extra dimensions are given by    
$h_{1,0}=r$ and $h_{2,0}=R$, where the zero subscript stands for the 
stabilized values. Therefore the stabilized volume is given by
$vol_0 = r^n R^p$, as stated in Eq.~(\ref{add1}).

With the help of Eqs.~(\ref{hij}) and (\ref{rg2}) we obtain the non diagonal
 kinetic terms to be
 \be
 {1\over 2\kappa^2}\left\{ n(n+2)\,\left({\partial_\mu h_1\over h_1}\right)^2 
 + 2\, np\, \left( {\partial_\mu h_1 \over h_1}\right)\, 
  \left( {\partial^\mu h_2\over h_2}\right)
 + p(p+2)\,\left({\partial_\mu h_2\over h_2}\right)^2
 \right \}\,.
 \label{kt}
 \ee
We diagonalize this by introducing two physical radion fields
 \be
 \sigma_v = M_{p}\sqrt{{n+p+2\over 2(n+p)}}~ 
 \ln \left({\sqrt{h}\over vol_0}\right)\,,
 \qquad \qquad
 \sigma_\bot =
 {M_{p}}\,\sqrt{{np\over n+p}}~\ln \left({h_2\over h_1}\, {r\over R}\right)\,.
  \label{radions}
 \ee
This makes the kinetic terms in Eq.~(\ref{kt}) diagonal and
canonically normalized. Both the radions have a straightforward 
physical interpretation, $\sigma_v$ is a sole function of $\sqrt{h}$, 
therefore it is related to the variation of the physical size of the 
volume, and often called a volume modulus (or volume radion). 
On the other hand 
$\sigma_\bot$ is related to the ratio of the sizes of both large 
and small dimensions, and we will often call it the shape  mode. 
Furthermore, note that both the radion fields are set 
to be zero when stabilized (that is when $h_1$ and $h_2$ obtain the values
$r$ and $R$ respectively).

The effective four dimensional action for the gravity sector can then be 
written as
 \be
 S=  \int\!d^4x\,\sqrt{-g}\, \left \{
    { 1\over \kappa^2}\,R[g] + {1\over 2} (\partial_\mu \sigma_v)^2
    + {1\over 2} (\partial_\mu \sigma_\bot)^2
    - U(\sigma_v,\sigma_\bot)
 \right \}\, .
 \label{gaction}
 \ee
Note that in a particular case when $n=0$, $\sigma_\bot =0$ and  we are
reduced to a single radion scenario of large extra dimensions.
The last term in the above action represents the
radion potential that must be introduced by hand to  
stabilize the extra dimensions. Its origin is
largely unknown so far. In this paper we will only assume that
there exists such a potential $U(\sigma_{v},\sigma_{\bot})$ and consider
illustrative forms for it in our discussion.
As we will discuss later on,  the detailed form of
$U(\sigma_{v},\sigma_{\bot})$  is important
in understanding the dynamics of the radions during and after inflation.

\section{Radion couplings}

Given the above definitions for the radions, we now discuss how they
couple to matter, both in the bulk and on the brane. We start by
considering a bulk scalar field, $\tilde\phi$ which latter on 
could be identified as the inflaton. The action, before performing the
conformal transformation on the metric, goes as
 \be
 S_{\tilde\phi}~=~\int\, d^4x\, d^{n+p}y\, \sqrt{-g}\sqrt{h}
  \left[{1\over 2}\, G^{AB}\partial_A\tilde\phi\, \partial_B\tilde\phi
   - V_{bulk}(\tilde\phi)\right]\,.
 \ee
We now take only the zero mode level of the theory, therefore we write
$\tilde\phi(x,y) = \phi(x)/\sqrt{vol_0}$, and properly scale the
potential by introducing an effective four dimensional potential,
$V(\phi)~=~vol_0\cdot V_{bulk}(\tilde\phi)$. By integrating out the
extra dimensions and introducing the conformal transformation, we
obtain an effective scalar action in the Einstein frame,
 \be
 S_\phi~=~\int\, d^4x\,\sqrt{-g} ~e^{2\varphi}{\sqrt{h}\over vol_0}\,
  \left[{1\over 2}\, g^{\mu\nu}\partial_\mu\phi\, \partial_\nu\phi
   - e^{2\varphi}V(\phi)\right] ~.
 \ee
It is now straightforward to see that the last term gives rise to a
coupling of $\phi$ to $\sigma_v$ due to the definition of $\varphi$
given in Eq.~(\ref{ephih}). This will also take care of the global
exponential term, which cancels the $\sqrt{h}/vol_0$ piece of the action.
Therefore we can simply write 
\be
 S_\phi~=~\int\, d^4x\,\sqrt{-g}
  \left[{1\over 2}\, g^{\mu\nu}\partial_\mu\phi\, \partial_\nu\phi
   - e^{-\alpha {\sigma_v / M_{p} }}V(\phi)\right]\,,
 \label{sphi}
 \ee
where the coupling constant is given by
\begin{equation}
\alpha = \sqrt{\frac{2(n+p)}{n+p+2}}\,.
\end{equation}

It is worth stressing  that the $\sigma_\bot$ radion does not couple
to the bulk scalar field. Furthermore, since the radion couplings
appear due to the conformal transformation in Eq.~(\ref{conformal}),
which depends only on $\sigma_v$, we can conclude the same is actually
true for any bulk field. Therefore the $\sigma_\bot$ radion will only 
couple to the brane matter, which we shall discuss now. Note that at 
this level we already notice an important difference in the dynamics 
of the two radion fields.

Let us now consider a brane scalar field, $\tilde\chi$. The general
action would be given as in $4+n$ dimensions, where the SM fields
propagate. This means that we have already integrated out extra $p$ large
dimensions which are harmless for the brane fields. Once more we stay
at the zero mode level, we introduce a properly normalized zero
mode by defining  $\tilde\chi =\chi/\sqrt{r^n}$. We start with the 
effective action 
 \be
 S_\chi~=~\int\, d^4x \sqrt{-g}\,\left({h_1\over r }\right)^n\,
  \left[{1\over 2}\, g^{\mu\nu}\partial_\mu\chi\, \partial_\nu\chi
   - V(\chi)\right] ~.
 \ee
Notice that unlike the bulk field case, the induced metric does not
contain the whole $\sqrt{h}$, but only the piece that goes along the
extra dimensions of the brane, $h_1^n$. Therefore after conformally
transforming the metric there will be a remnant of the $h$ metric.
Indeed by writing $h_1^n = \sqrt{h}~h_2^{-p}$ and then, using the
radion definitions in Eq.~(\ref{radions}) in order to replace the $h_2$ 
term, we obtain
 \be
\label{braneact}
 S_\chi~=~\int\, d^4x \sqrt{-g}
 ~e^{-\beta \sigma_v/M_{p}} ~ e^{-\gamma \sigma_\bot/M_{p}}~
  \left[{1\over 2}\, g^{\mu\nu}\partial_\mu\chi\, \partial_\nu\chi
   - e^{-\alpha \sigma_v/M_{p}}~V(\chi)\right]\,.
 \ee
where the coupling constants are given by
\begin{eqnarray}
\beta &=& \frac{ p}{\sqrt{(n+p)(n+p+2)}}\,, \\
\gamma &=& \sqrt{\frac{np}{n+p}}\,.
\end{eqnarray}
We now see that the $\sigma_\bot$ radion has a global coupling to 
both kinetic and potential terms of the brane fields, whereas 
extra global coupling  of $\sigma_v$ to both kinetic and 
potential terms also appear. Such couplings
affect the definition of the canonical brane field, which  are, however,
nicely recovered around the minima of $\sigma_{v,\bot}$.

As far as the other fields go, following a similar procedure, we find that
the radion couplings to brane matter are given as:
 \be
S_{gauge}~=~-\int\, d^4x \sqrt{-g}
 ~e^{(\alpha-\beta) \sigma_v/M_{p}}~e^{-\gamma\sigma_\bot/M_{p}}~
 {1\over 4}\, F_{\mu\nu}\, F_{\rho\tau}\, g^{\mu\rho}\, g^{\nu\tau}\,,
 \label{rgauge}
 \ee
for the zero mode gauge fields, and by
 \be S_{\psi}~=~\int\, d^4x \sqrt{-g}
 ~e^{-(\alpha/2+\beta)\sigma_v/M_{p}}~e^{-\gamma\sigma_\bot/M_{p}}~
 \left( -i \bar{\psi}D_\mu \gamma_a \psi \right)~e^a_\nu~g^{\mu\nu}\,,
 \ee
for the zero mode fermions. It is worth mentioning that if we take $n=0$, 
as for the case of large extra dimensions, we find that $\alpha = \beta$,  
and thus the  radion coupling to a massless gauge field 
vanishes~\cite{lykken}. This is however not the case when the matter 
fields propagate only in part of the extra space as shown in
Eq.~(\ref{rgauge}).

\section{Initial conditions before inflation}

The most natural scale for the universe to start in $4+d$ dimensions 
is the fundamental scale. Therefore, by taking $h_{1,2}(t=0)\sim M_{\ast}^{-1}$
as the initial values for the size of the extra dimensions, we obtain 
the initial vacuum expectation values (vevs) for $\sigma_{v}$ and 
$\sigma_{\bot}$ are of the order of the Planck scale:
 \be
  \label{init}
  \sigma_v(0)~=~-{2}\,M_{p}\sqrt{{n+p+2\over n+p}}
   \ln\left({M_{p}\over M_{\ast}} \right)\,, 
  \qquad \sigma_\bot(0)~=~
   - M_{p}\,\sqrt{{ np\over n+p}}~\ln\left(\frac{R}{r} \right)\,.
 \ee
Note that both the initial values  for the radions are negative. 
This is due to the fact that the  extra dimensions are relatively 
larger today (and so the volume). Physically, the extra dimensions 
have to grow, while both the radions approach  to zero from negative values.  
For the sake of illustration we can take $M_{\ast}\sim {\cal O}(\rm TeV)$, 
and we find that indeed both the absolute values for the vevs, though
$n$ and $p$ dependent, will have initial vevs quite large, e.g. 
$\sigma_{v}(0)\sim {\cal O}(10 M_{p}),~\sigma_{\bot}(0)\sim{\cal O}(10 M_{p})$.

There is a  couple of interesting points to be mentioned. First,
the observed Universe is many orders of magnitude larger than 
$M_{\ast}^{-1}$, and it is also certainly larger than even the  
largest $p$ extra dimensions, which should be smaller than few 
micrometers in order to evade detection at small distance gravity 
experiments. Therefore an {\it asymmetric expansion} of the spatial 
dimensions is required. Besides the galaxy formation requires 
that there has to be perturbations seeded on large scales (larger 
than the size of the horizon). Inflation is the only paradigm 
which makes the universe large, flat, isotropic and 
homogeneous, if it began at $M_{\ast}^{-1}$. Even though it creates a
homogeneous and isotropic universe, it cannot really beat the quantum
fluctuations which are being stretched outside the horizon at the 
time of inflation. These quantum fluctuations when they reenter into
our horizon late during the radiation dominated era provide the seed
for the galaxy formation~\cite{Brandenberger}.

However what remains to be seen is what could be the inflaton in our 
scenario. We should also remind the reader that at the moment fundamental
theory provides no guidance regarding the radion potential and, therefore 
the masses for $\sigma_{v},~\sigma_{\bot}$ are not clearly fixed by our action
Eq.~(\ref{gaction}). This leads to a degree of speculation 
which we will discuss briefly.

\section{Radion mass}

Although, from a naive point of view one would expect that the radion
masses ought to be somehow related to the compactification scale, this
seems 
to be less obvious when we recognize that physical radions in an 
asymmetric bulk are not solely connected to a particular
extra dimension, but a combination of all. 
Moreover, since we have no knowledge about the origin and actual form 
of the stabilization potential, very little can be said about the radion
masses without further assumptions. It has been conjectured that the
radion mass can
be as small as $m_r\ls M_c^2/M_{p}$~\cite{chacko} for $M_c = 1/r$. 
This seems certainly
true for the case where the stabilization potential comes from brane
physics, however the mass can be much larger if bulk physics plays a role
as we argue below.

For the case where the stabilization potential is based on bulk
physics there is a volume enhancement in the effective four
dimensional theory. On pure dimensional grounds, we can see that 
the potential could be as large as $M_{\ast}^2M_{p}^2$, rather 
than $M_{\ast}^4$ leading to much larger radion masses. For instance, in 
this case, we could write the radion potential 
roughly as $U(\sigma)\ls M_{\ast}^2M_{p}^2 U(\sigma/M_{p})$, so that 
the upper bound on the radion mass comes naturally as $m_r^2\ls M_{\ast}^2$. 
Another intuitive way of seeing this result is by noticing that masses 
for the radions $m_{r~i}^2\sigma_{i}^2$, where $i=v,\bot$, are 
invariant under dimensional reduction. The possibility of having a 
large radion mass has also been pointed out in another example 
given in Ref.~\cite{nima2}, where radion mass and compactification 
scale are exponentially related to each other, therefore reconciling  
the likelihood of large dimensions with large radion masses.

As it has already been noticed~\cite{chacko} a light mass radion can 
mediate long range interactions with a similar strength as gravity, therefore, 
the experiments testing the Newton's law at small distances~\cite{expgrav} 
impose a lower bound on such radion masses, which corresponds to the mass 
scale, $m_r>10^{-3}$~eV.

\subsection{A phenomenological guess on radion potential}

While we have no guidance from a fundamental theory regarding the nature
of the radion potential, there are some cosmological constraints which
could be used to guess its overall nature. First of all, we would like to
have the minimum of the radion potential at $\sigma_v=0$ and
$\sigma_{\bot}=0$. Secondly, since the volume
modulus, $\sigma_v$ governs the magnitude of the Newton's constant, it
must be stabilized before BBN, in order to be
compatible with stringent bounds on the variation of the Newton's constant
during BBN~\cite{clifford}. The shape mode must also be stabilized 
in order to evade constraints from long range interaction between the 
SM fields and late time variations of coupling constants. Since 
the stabilization crucially depends on the actual form of the 
stabilization potential, this provides some constraint on its form.

The two extreme possibilities for the potentials are as
follows:

\begin{itemize}

\item[(i)] The potential for the radions could be fairly flat, even 
for larger vevs, in which case a simple mass squared-like potential,
e.g. $U\sim {1\over 2} m_r^2\sigma_r^2$, where $r=v,~\bot$ can be used as 
a phenomenological model. As we will show this choice of potential
can be extremely hard to stabilize. Besides the late oscillations
of the radions will give rise to a moduli problem, for which there is
no simple solution.

\item[(ii)] This problem can be ameliorated provided that the moduli 
potential is steeper than $m_{i}^2\sigma_{i}^2$ at large vevs, or 
in other words, the flatness of the potential is lifted. Let us consider 
for example a phenomenological potential that can give rise 
to a volume stabilization
 \be
 U(vol)\sim \mu^2 M_{\ast}^{2+2(n+p)} \left(vol_{0}-vol_{phys}\right)^2\,,
 \ee
where the overall scale has been taken in the general philosophy of using 
$M_\ast$ to adjust the dimensions of the potential. Here $\mu$ is assumed 
to be some mass scale naturally related to the volume, and so this is 
at most of order $M_\ast^2/M_P$. By using  Eq.~(\ref{add1}), we can 
rewrite the potential as
 \be
 U(vol) =
  m_{v}^2 M_{p}^2 \left(1-vol_{phys}/vol_0\right)^2
  = m_{v}^2 M_{p}^2 \left(1-\sqrt{h}/vol_0\right)^2\,,
  \ee
where now the mass scale is $m_v = \mu M_P/M_\ast$, which is much larger than
$\mu$, but yet smaller than $M_\ast$. The above equation  
translates into a potential for the volume modulus with
an exponential profile:
 \be
\label{guess}
U(\sigma_v)\sim m^2_{v}M_{p}^2\left(1-e^{\alpha_v\sigma_{v}/M_{p}}\right)^2\,.
 \ee
Furthermore,  note that the
radion mass around the minimum 
becomes $m_{v}$, therefore reinstating our arguments.
Some interesting facts about the above potential are that (i) 
it is exponentially  steep for $\sigma_v >0$, 
which means that such a 
potential will not allow the volume to grow much larger than its stable size. 
(ii) On the other hand, the potential is fairly flat for large and
negative 
vevs of $\sigma_v$, allowing the volume to easily expand. 
Let us stress here that we are simply using this potential  
for illustrative purposes.

\end{itemize}

Needless to mention that the above arguments should also hold 
for the shape mode. However unlike the volume modulus it 
is extremely hard to make any speculation on the mass for the 
shape mode. The only obvious constraint comes from the long
range interaction mediated by $\sigma_{\bot}$, which gives
the lower bound $m_{\bot}\geq 10^{-3}$~eV, as already mentioned. 
In summary we can safely argue that the mass range for both the 
radions could be
\be
\label{range}
10^{-3}~{\rm eV}\leq ~m_r~\leq M_{\ast}\,,
\ee
although there are models where the mass of the volume modulus radion
could easily be close to the upper bound.
\section{Decay rates of the two radions}

Massive radions may decay into visible particles. Decay channels may vary 
depending on the actual mass. However, without loss of generality we can say
that radions decay into photons, and possibly other SM particles. 
The couplings that are responsible for 
such processes are described in section three. In all cases, but when 
the SM fields are truly  four dimensional (when $n=0$), the decay would 
take place at tree level and, when evaluated around the stabilized radions
they will always come with Planck suppressed couplings. The decay rate 
for any channel then goes as
 \be 
   \Gamma_{r} \approx a^2 {m_r^3\over 192 \pi M_{P}^2}\, ,
 \label{gammar}
 \ee
where the coupling constant $a\sim {\cal O}(1)$  depends on the 
type of radion, specific decay channel and the number of
dimensions. $a$ is actually given in terms of the previously 
introduced coupling constants $\alpha$, $\beta$ and $\gamma$ 
(see section three). The life time of the radion is then given by
 \be 
   \tau_{r} \approx 
   1.2 \times 10^{7}\, {\rm yr} ~\left( {{\rm GeV}\over m_r}\right)^3~.
 \label{taur}
 \ee
It is illustrative to see that  a radion of mass $m_r\approx 30$~MeV 
has a life time $\tau_r\approx 10^{11}$~yr, which is already about the 
age of the universe. Lighter radions would then be effectively stable,
so they might be good candidates for the cold dark matter. For heavier 
masses the radions would decay into visible channels (photons and leptons) 
which might leave a detectable trace. For instance, for masses just around 
few MeV, radions can be produced not only in the early universe but also in 
supernova cores. Its later decay would inject hard photons of MeV 
energy at present time contributing  
to the diffuse gamma ray background flux, however, it is not too 
different from the one induced by KK graviton decays~\cite{hannestad}, 
and so, no new bounds on the fundamental scale can be obtained from 
this effect. In contrast, a radion mass about a few TeV would decay
into energetic SM particles just right at the BBN era ($\sim$~1-10~sec),
which may photo-disintegrate  primordial nuclei. 
Obviously, a radion mass within  MeVs up to just below TeVs  
should decay somewhere within BBN and present time. 
Radions with larger masses than 10 TeV, or so, 
would decay fast enough as to wash out any primordial radion abundance.
They could potentially reheat the universe with the SM degrees of freedom.

\section{Radion Driven Inflation}

The two radions evolve dynamically. It could be interesting to check 
whether by choosing an appropriate radion potential, one can drive
inflation or not. Let us 
consider type I potentials for $\sigma_{v},~\sigma_{\bot}$. As already
mentioned, irrespective of the masses $m_{v},~m_{\bot}$, the initial 
vevs are large in $4$ dimensions, e.g. 
$|\sigma_{v}(0)| \sim |\sigma_{\bot}(0)| \sim {\cal O}(10~M_{p})$.
At such large vevs the initial conditions are equivalent to chaotic 
initial conditions, and if there is no other source for cosmological
constant, then inflation should take place along the line of 
double inflation~\cite{polarski} 
\be
V\sim {1\over 2}m^2_{v}\sigma^2_{v}+ {1\over 2}m^2_{\bot}\sigma^2_{\bot}\,.
\ee 
Both the components give rise to the Hubble expansion, for which we obtain
$H\approx 10~{\rm max}[m_v, m_\bot]$. Using the mass window in
Eq.(\ref{range}), we see  that $H_{inf}$  could lie between 
${\cal O}(10^{-2}~{\rm eV})\leq H_{inf}\leq {\cal O}( M_{\ast})$. 
Inflation will terminate
when slow roll conditions are no longer valid, which happens in
the chaotic model when $\sigma_{v},~\sigma_{\bot}\sim \sqrt{2}M_{p}$.

During inflation both the fields generate quantum fluctuations 
which leave the horizon. We can estimate the amplitude for their
density perturbations as
\begin{equation}
\frac{\delta \rho}{\rho} \approx 
   \frac{{\rm max}\left[m_v,m_\bot\right]}{M_{p}}< {M_\ast\over M_P}\,.
\end{equation} 
Note that the present constraint on the amplitude for the primordial
density perturbations is given by one part in $10^{5}$, this is the COBE 
normalization verified by recent experiment such as WMAP~\cite{WMAP1,WMAP}.
Unfortunately neither $m_{v}$ nor $m_{\bot}$ is sufficiently large 
to generate the right amplitude for the density perturbations in this
radions driven inflation, as the bound on the RHS of the last equation
indicates.

An interesting possibility may arise if the total radion potential
is given by a combination, such as
\begin{equation}
V\sim m_{v}^2M_{p}^2\left(1-e^{\alpha_v\sigma_{v}/M_{p}}\right)^2 +
{1\over 2} m_{\bot}^2\sigma_{\bot}^2\,,
\end{equation}
where we assume a particular scenario with $m_{v}\sim{\cal O}(1~{\rm TeV})$,
assuming for instance that there is some bulk physics which gives rise to
the origin of volume stabilization, for e.g. flux quantization~\cite{Frey}.
Let us also assume for the time being $m_{\bot}\ll m_v$. Note that here we get
inflation from $\sigma_{v}$ radion, because the potential has a 
flat direction for large and negative vevs, e.g. 
$\sigma_{v}(0)\sim-{\cal O}(10 M_{p})$. Clearly for this case 
the exponential part on the potential is negligible, and the 
cosmological constant dominates, such that $V\sim m_v^2M_{P}^2$.
The Hubble expansion during inflation is given by $H_{inf}\approx m_v$.
Again it is easy to verify that the adiabatic density perturbations produced 
by $\sigma_{v}$ during inflation will be way too small compared to
$10^{-5}$, therefore one has to seek alternative ways of generating
density perturbations. One such possibility is described below.

Whenever there are more than one fields involved during inflation,
the quantum fluctuations can be of two types, pure adiabatic fluctuations,
which tells us about the perturbations along the final trajectory of the
fields, and second one is the isocurvature fluctuations which designates the
perturbations along the orthogonal direction of the trajectory. In general
isocurvature fluctuations arise due to difference in pressure fluctuations
and most importantly isocurvature fluctuations feed the adiabatic 
fluctuations outside the horizon.

In our case $\sigma_{\bot}$ being light compared to the Hubble 
expansion during inflation will also have homogeneous fluctuations. 
These fluctuations are actually isocurvature in nature. 

The amplitude for the isocurvature fluctuations generated by 
$\sigma_{\bot}$ during inflation can be given by~\cite{Iso}
\be
\label{iso}
{\cal P}^{1/2}_{\sigma_{\bot}}\sim \frac{H}{2\pi\sigma_{\bot}}\sim
\frac{m_{v}}{2\pi\sigma_{\bot}}\,.
\ee
In order to obtain the right amplitude for the density perturbations,
$\sigma_{\bot}\sim 10^{5}\times m_v$. If $m_v\sim 1$~TeV,
then the vev of $\sigma_{\bot} \sim 10^{8}$~GeV. However since the mass of 
$\sigma_{\bot}$ is very light and the initial amplitude, i.e. 
$|\sigma_{\bot}|\sim {\cal O}(10 M_{p})$ is quite large, it will be extremely
hard for $\sigma_{\bot}$ to roll down 
from $M_{p}\sim 10^{18}$~GeV to the amplitude $10^{8}$~GeV. 
Especially note that during inflation  the 
radions are both slow rolling. Therefore the evolution of 
$\sigma_{\bot}$ can be written as
\begin{equation}
3H\dot\sigma_{\bot}\approx -m^2_{\bot}\sigma_{\bot}\,.
\end{equation}
Integrating this out we obtain
\begin{equation}
\label{eqmot}
\sigma_{\bot}(t)\approx \sigma_{\bot}(0)e^{-(m^2_{\bot}/3H)t}\,.
\end{equation}
For $m_{\bot}\ll H$, it would take many e-foldings of inflation
before the field could be settled down to its minimum. Anther way 
of saying this is that in order to obtain $\sigma_{\bot}\sim 10^{8}$~GeV
from the initial value $10~ M_{p}$, it would require between
$76\leq {\cal N}_{e} \leq 10^{16}$, for 
$1~{\rm TeV}\geq m_{\bot}\geq 10^{-3}$~eV, respectively, 
where ${\cal N}_{e}$ is the number of e-foldings. 
There is certainly a possibility that one of the radions could
drive inflation while the other generates adequate density perturbations. 
If this can be realized then this will be the simplest model to be considered.


\subsection{Reheating $\&$ moduli problem}

Let us briefly discuss the reheating temperature due to the decay of
radions. During coherent oscillations, the radions can decay into SM
degrees of freedom.
The couplings among radions and brane SM fields have been discussed in 
section 3. By looking at those expressions it is easy to see that the 
presence of a non trivial radion vacuum would have two effects: 
it can change the normalization of the wave functions through the global
radion
couplings to the kinetic terms; and it can affect the gauge and 
Yukawa couplings as well as the potential terms. 
Nevertheless the overall radion couplings remain Planck suppressed.
Therefore for our naive estimations one may still use the results given 
in section 6.

Radion decay products must hadronize before BBN~\cite{Jaikumar}. If however 
the radion decays much later than BBN then there will be two problems:
(1) hadronizing the universe with SM relativistic degrees of freedom 
may not be complete, and (2) even if thermalization is completed well
before BBN, the radion decay products with hard momentum can eventually
destroy the light elements by reheating the universe.

Using Eq.~(\ref{gammar}) the perturbative reheating temperature can be 
estimated by
 \be 
 T_r\approx 0.1 \sqrt{M_P\Gamma_r} = 
 1.28\times 10^{-7}~{\rm GeV}~\cdot\left({m_v\over {\rm TeV}}\right)^{3/2} \,.
 \ee
Therefore, from our naive analysis, it seems that the 
only solution to this problem is that the radion $\sigma_{v}$ must
be heavier than about $500$~TeV. Indeed  for a mass at such  
value the reheating temperature is just about $T_r\sim 1$~MeV.
This also forbids the possibility of having a TeV fundamental scale
in this scenario\footnote{ Interesting possibility may arise if the
reheating 
temperature is asserted to be the Hagedorn temperature, which comes 
out to be close to the mass of the radions, i.e. 
$T_{r}\sim 0.74\times m_{v}\leq {\rm TeV}$, see~\cite{Jaikumar1}.}.

Let us discuss the associated moduli problem. In the above scenario 
a moduli problem will occur due to the coherent oscillations of 
$\sigma_{\bot}$. Note that $\sigma_{\bot}$ oscillates
with a frequency $m_{\bot}$ when $H\sim m_{\bot}$. Depending on 
the initial amplitude of the coherent oscillations of $\sigma_{\bot}$ 
there can be adverse affects on cosmology. The energy density stored in 
$\rho_{\sigma_{\bot}}\approx m^2_{\bot}\sigma^2_{\bot}$ will redshift 
as in a matter dominated era and will eventually dominate the relativistic
bath already created by the decay products of $\sigma_{v}$, provided
$\rho_{\sigma}\approx H^2 M_{p}^2$. However this requires the amplitude of
$\sigma_{\bot}$ oscillations to be as large as $\sigma_{\bot}(0)\sim M_{p}$. 
On the other hand, if the initial amplitude of the oscillations are
sufficiently low, e.g. $\sigma_{\bot}\leq M_{p}$ then it is certainly 
possible to avoid the radion dominance problem. This may occur if there 
is a large enough e-foldings of inflation.

\section{Brane Inflation and its consequences}

In this section we consider a possibility where the 
inflaton potential arises from brane physics. The scalar
field action is given by Eq.~(\ref{braneact}). We suppose that
a hypothetical scalar field $\chi$ is living on the $4+ n$ dimensional 
brane and its potential is given by $V(\chi)$. The total action with 
the radions is given by

\begin{eqnarray}
S=\int d^4x\sqrt{-g}\left[\frac{1}{\kappa^2}R+\frac{1}{2}(\partial_{\mu}
\sigma_{v})^2+\frac{1}{2}(\partial_{\mu}\sigma_{\bot})^2 + \frac{1}{2}
e^{-(\beta\sigma_{v} +\gamma\sigma_{\bot})/M_{p}}(\partial_{\mu}\chi)^2
\right. \nonumber \\
\left.- e^{-((\beta+\alpha)\sigma_{v}+\gamma\sigma_{\bot})/M_{p}}V(\chi)-
U(\sigma_{v}\sigma_{\bot})\right]\,.
\end{eqnarray}
The natural scale for the effective $V(\chi)$ in $4$  dimensions is
\be
V(\chi)\ls (M_*r)^nM_*^4\simeq 10^{n}M_{\ast}^4\,.
\ee
The numerical factor on the RHS arises 
when the compactification scale for $n$ extra 
spatial dimensions is $M_{c}\sim 10^{-1}M_{\ast}$. 
Initial phase of inflation is governed by the exponential part 
$\exp\left[-((\beta+\alpha)\sigma_{v}+\gamma\sigma_{\bot})/M_{p}\right]V(\chi)$,
since
$\sigma_{v}(0),~\sigma_{\bot}(0)\simeq -{\cal O}(10~M_{p})$, which exponentially
enhances that contribution.

In slow roll approximations, where we neglect the higher order
time derivatives and we assume that the friction term proportional
to the Hubble expansion dominates, the  equations of motion 
for the fields  and the Hubble parameter 
simplify a lot, and one gets
 \begin{eqnarray}
  3H\dot\sigma_{v} &\approx& \frac{\beta+\alpha}{M_{p}}
  e^{-((\beta+\alpha)\sigma_{v}+\gamma\sigma_{\bot})/M_{p}}V(\chi)\,, \\
  3H\dot\sigma_{\bot} &\approx&\frac{\gamma}{M_{p}}
  e^{-((\beta+\alpha)\sigma_{v}+\gamma\sigma_{\bot})/M_{p}}V(\chi)\,, \\
  3H\dot\chi &\approx&-e^{-(\alpha\sigma_{v})/M_{p}}\frac{\partial V}
  {\partial\chi}\,,\\
  H^2&\approx&\frac{1}{3M_{p}^2}\left[e^{-((\beta+\alpha)\sigma_{v}+
  \gamma\sigma_{\bot})/M_{p}}V(\chi)+U(\sigma_{v},\sigma_{\bot})\right]\,.
 \end{eqnarray}
 Let us assume that during inflation 
 $\exp\left[-((\beta+\alpha)\sigma_{v} 
+\gamma\sigma_{\bot})/M_{p}\right]V(\chi)\gg U(\sigma_{v},\sigma_{\bot})$.
We can also assume that $V(\chi)$ piece is dominated by a false vacuum
and it is merely constant during inflation. However, note that
$\sigma_{v}$ and $\sigma_{\bot}$ are present in the exponential part of
the potential. The rolling of the radions will govern the initial phase 
of inflation in such a scenario. This is an example of assisted inflation, 
for which the scale factor follows~\cite{assist}
 \be
 a(t)\propto t^{\kappa}\,,~\quad \kappa =\frac{2}{(\alpha+\beta)^2+\gamma^2}\,.
 \ee
Here we have assumed that $V(\chi)$ is a flat potential, which means that
$\chi$ evolution is slower compared to $\sigma_{v},~\sigma_{\bot}$. In our
case the power law inflation lasts as long as 
$|\sigma_{v}|,~|\sigma_{\bot}|\sim M_{p}$.
Once the vevs of both the radions become smaller than $M_{p}$, then
effective masses for the radions become
\begin{eqnarray}
\label{effecm1}
m^2_{v~eff} &\approx& \frac{\partial^2 U}{\partial\sigma_{v}^2}+
(\beta+\alpha)^2\frac{V(\chi)}{M_{p}^2}\,, \\
\label{effecm2}
m^2_{\bot~eff} &\approx&\frac{\partial^2 U}{\partial\sigma_{\bot}^2}+
\gamma^2\frac{V(\chi)}{M_{p}^2}\,.
\end{eqnarray}
If  $V(\chi)/M_{p}^2$ were greater than the bare masses for
the radions, then an effective radion mass would become of order
$H_{inf}\sim V(\chi)/M_{p}^2$. However, since 
$V(\chi)\ls (M_\ast r)^n~M_{\ast}^4$, this means $H_{inf}$ is 
many orders of magnitude smaller than $M_*$. For instance, for 
$M_{\ast}\sim 1$~TeV and  $M_*r\sim 10$, we obtain an inflationary 
scale of  order $H_{inf}\ls 10^n\times 10^{-2}$~eV!. 
This also implies that the radions potential, $U$, should be extremely flat.
Certainly $H_{inf}$ improves for larger vales of $n$ and 
$M_{\ast}$, but still it is hard to push the scale of 
inflation up to ${\cal O}(\rm TeV)$. Such low scale of inflation 
has many problems to produce successful baryogenesis
and BBN~\cite{Manyinf}.

On the other hand if radion masses win over inflaton scale in 
Eqs.~(\ref{effecm1}) and (\ref{effecm2}), then inflation ends right after
$\sigma_{v},~\sigma_{\bot}\sim M_{p}$. The radions start oscillations 
around their minimum and eventually decay into SM degrees of freedom.
Note that the radion oscillations are large with the amplitudes of the order 
of $M_{p}$. The late domination of radion can occur
if there is a hierarchy in radion masses. The lighter 
one, if it does not decay completely, can cause a problem 
similar to the moduli problem as discussed before.

\section{Bulk inflation}

As a final possibility we investigate the bulk driven inflation.
We assume that inflation  arises from the bulk physics,
e.g. a scalar field $\phi$ living in $4+d$ dimensions, such that the 
action is given by
\be
S=\int d^4x \sqrt{-g}\left[\frac{1}{2}(\partial_{\mu}\sigma_{v})^2+
\frac{1}{2}(\partial_{\mu}\sigma_{\bot})^2+\frac{1}{2}(\partial_{\mu}\phi)^2
-e^{-\alpha\sigma_{v}/M_{p}}V(\phi) - U(\sigma_{v},~\sigma_{\bot})\right]\,.
\ee
The four dimensional action is much simpler in this case. Note that after
dimensional reduction the effective potential is 
$V(\phi)\ls M_{\ast}^2M_{p}^2$, which can be relatively large compared to the 
former scenario where we assumed that inflaton potential as originating
from the brane physics. Let us discuss this scenario in some detail.

First of all note that the radion $\sigma_{v}$ only couples to the inflaton
potential and not $\sigma_{\bot}$. Therefore $\sigma_{\bot}$ does not 
get any effective mass correction through the inflaton potential, 
while the volume modulus does. As we shall see this will play an interesting
role in ameliorating the moduli, or late domination of radion in thermal
history of the universe.  

\subsection{Inflationary dynamics}

The equations of motion for the above action are given by
\begin{eqnarray}
&&\ddot\sigma_{v}+3H\dot\sigma_{v}+\frac{\alpha}{M_{p}}
e^{\alpha\sigma_{v}/M_{p}}V(\phi)-\frac{\partial U}{\partial\sigma_{v}}=0\,,\\
&&\ddot\sigma_{\bot}+3H\dot\sigma_{\bot}-\frac{\partial U}
{\partial \sigma_{\bot}}=0\,, \\
&&\ddot\phi+3H\dot\phi -e^{-\alpha\sigma_{v}/M_{p}}\frac{\partial V}
{\partial \phi}=0\,,\\
&&H^2=\frac{1}{3M_{p}^2}\left[\frac{(\dot\sigma_{v})^2}{2}+
\frac{(\dot\sigma_{\bot})^2}{2}+\frac{(\dot\phi)^2}{2}+
e^{-\alpha\sigma_{v}/M_{p}}V(\phi)+ U \right]\,.
\end{eqnarray}
In this particular scenario there are two phases of inflation possible.
Initial phase of inflation is driven by the exponential part of the 
potential provided there is flat potential for $V(\phi)$. This phase is 
essentially driven by radion, and the subsequent phase
of inflation is governed by $V(\phi)$ with a Hubble parameter
$H\ls M_{\ast}$. During inflation the slow roll equations are given to a  
good approximation by
\begin{eqnarray}
3H\dot\sigma_{v}&\approx& -\frac{\alpha}{M_{p}}
e^{\alpha\sigma_{v}/M_{p}}V(\phi)+\frac{\partial U}{\partial \sigma_{v}}\,, \\
3H\dot\sigma_{\bot}&\approx &\frac{\partial U}{\partial \sigma_{v}}\,, \\
3H\dot\phi &\approx & e^{-\alpha\sigma_{v}/M_{p}}\frac{\partial V}
{\partial \phi}\,, \\
 H^2&\approx &
 \frac{1}{3M_{p}^2}\left[e^{-\alpha\sigma_{v}/M_{p}}V(\phi) +U \right]\,.
 \label{bulkexp} 
\end{eqnarray}
The scale factor $a(t)$ follows
\be
a(t)\propto t^{(n+p+2)/n+p}\,,
\ee
by assuming that $V(\phi)\gg U$ and $\phi$ evolution is extremely
slow.  Note that again we obtain a power law inflation. The power law
inflation does not last for ever. Once $\sigma_{v}\sim -M_{p}$,
inflation is not supported along $\sigma_{v}$ direction, rather
inflation is supported by the inflaton potential $V(\phi)$. Also note
that when the vev of $\sigma_{v}$ becomes less than $M_{p}$, we can
expand the exponential term $\exp({-\alpha\sigma_{v}/M_{p}})$.  The
effective mass for $\sigma_{v}$ can be expressed as
\be
m^2_{\sigma_{v}~eff}\approx \alpha \frac{V(\chi)}{M_{p}^2}+
\frac{\partial^2 U}{\partial \sigma_{v}^2}\simeq 3\alpha H_{inf}^2\,. 
\ee
Where we have assumed $H_{inf}\approx V(\phi)/M_{p}^2$, irrespective of
the form of $V(\phi)$, provided of course
that $V(\phi)$ is sufficiently flat to 
support inflation.  Further note that if the bare mass for $\sigma_{v}$ 
is small compared to $H_{inf}\sim M_{\ast}$, then the Hubble induced mass
term will dominate. Once Hubble induced mass term is switched on the radion
$\sigma_{v}$ will follow the evolution
\be
\sigma_{v}(t)\sim \sigma_{v~i}~e^{-(m^2_{eff}/3H)t}\,,
\ee
where $\sigma_{v~i}$ is the initial amplitude which is of order $M_{p}$.
The volume modulus therefore settles in its local minimum within one Hubble
time. 
Note, however, that the  global minimum for $\sigma_{v}$ could be displaced 
during inflation. If such a displacement is large enough another phase of
inflation could take place. That generally happens when the radion potential 
is quite flat~\cite{kolb}. We will discuss this issue in the next subsection.

Further note that although the amplitude of $\sigma_{v}$ can be 
reduced considerably during inflation, the same can not be true 
for the shape mode, $\sigma_{\bot}$. The other radion does not
obtain a Hubble induced mass correction during inflation, because 
it does not couple to $\phi$ field at all at classical level.
Therefore the amplitude of $\sigma_{\bot}$ may not damp as much as we
wish. Depending on its mass and its vev it might be possible that
$\sigma_{\bot}$ could be responsible for generating adiabatic density
perturbations, which we will discuss later on.

As noted earlier, we could easily get two subsequent bouts 
of inflation; initially driven by the radion $\sigma_{v}$
and then by the inflaton field in the bulk. However number
of e-foldings required for the structure formation depends on
the exact thermal history of the universe. Usually the number of 
e-foldings is much less than $60$ if the scale of inflation is low.
Especially if the scale of inflation is as low as 
$H_{inf}\approx M_{\ast}$, it was already shown in 
\cite{Lorenzana1,Green} that the number of e-foldings required for 
structure formation would be $43$, provided the universe reheats 
by $T_{rh}\sim 10-100$~MeV. Therefore it is only the last $43$ e-foldings of
inflation which are important for the purpose of density perturbations.

So far we have not specified the form of $V(\phi)$. However we have 
noticed that the upper bound on bulk inflation energy density is
$\sim M_*^2M_{\rm p}^2$. Therefore the second bout of inflation,
which is mainly driven by the bulk field, $\phi$, the scale of
inflation, $H_{inf}\sim M_{*}$, which for $M_{\ast}\sim {\cal O}(\rm TeV)$,
is sufficiently low. It is easy to demonstrate that a single  bulk inflaton
model does not produce adequate density 
perturbations (see for instance the discussions in Ref.~\cite{Manyinf}),
whereas  a two field bulk inflation
model, where inflation ends via a phase transition similar to hybrid inflation
scenario, can generate the COBE normalization~\cite{Mohapatra}.

The effective four dimensional potential for two fields is given by
\cite{Mohapatra}
\begin{equation}
\label{hybp0}
V(\phi, N) \equiv \left(\frac{M_{\rm p}}{M_{*}}\right)^2\lambda^2 N_0^4
+\frac{\lambda^2}{4}\left(\frac{M_{\ast}}{M_{\rm p}}\right)^2N^4-
\lambda^2 N_0^2N^2
+g^2\left(\frac{M_{\ast}}{M_{\rm p}}\right)^2\phi^2N^2 +
\frac{1}{2}m_{\phi}^2\phi^2\,,
\end{equation}
where $\phi,~N$ are zero modes of the bulk fields. Note that the four
dimensional
couplings are Planck suppressed, which comes out naturally by integrating
out the extra $d$ dimensions. The naturalness condition requires the bulk
couplings $\lambda,~g\sim {\cal O}(1)$, and $N_{0}\sim M_{\ast}$. Inflation
in this scenario is supported by the false vacuum $\lambda^2M_{\ast}^2M_{p}^2$,
because during inflation $N$ settles in its local minimum, $N=0$, 
and the rolling
of $\phi$ field gives rise to the phase transition. However one has to ensure
that the false vacuum contribution dominates over $(1/2)m_{\phi}^2\phi^2$
term. In this respect it is plausible to give a small mass term for the 
$\phi$ field, $m_{\phi}\ll M_{\ast}$. Note that in this scenario 
inflation ends when 
\begin{equation}
\phi_{c}= \frac{\lambda}{g}\left(\frac{N_0 M_{\rm p}}{M_{\ast}}\right)\,.
\end{equation}
Note, if 
$\lambda \sim g $, and, $N_0 \sim M_{\ast}$, we automatically get 
$\phi_{\rm c} \sim M_{\rm p}$.

This potential was further studied numerically in Ref.~\cite{Green}, and 
it was shown that in this scenario the phase transition which ends 
inflation happens extremely slowly. In the most natural case,
$N_{0}\sim M_{\ast}$, $\lambda,~g\sim {\cal O}(1)$, and 
$m_{\phi}\leq 0.1M_{\ast}$, it is possible to have extremely 
large number of e-foldings of order, ${\cal N}_{e}\sim {\cal O}(10^{6})$.

The occurrence of large number of e-foldings can be understood due to 
the fact that the shape of the potential along the $N$ direction is 
extremely flat due to extremely small coupling.
The field $N$ rolls down extremely slowly and spends a considerable time
near the top of the potential before rolling down to its global minimum
$N\sim \sqrt{2}M_{\ast}$, while the $\phi$ field rolls slowly away from 
$\phi_{c}$ towards its global minimum $\phi=0$. It can be shown that 
inflation ends in this picture when~\cite{Green} 
\begin{equation}
2 g^{2}\left(\frac{M_{*}}{M_{\rm P}}\right)^{2}N^{2}(t)\sim m_{\phi}^{2} \,,
\end{equation}
The phase transition could be made rapid, or the number of e-foldings 
during the phase transition can be made small, provided 
there is a hierarchy $m_{\phi}<< N_{0}< M_{\ast}$.

\subsection{The Radion Problem }

Let us consider once more the potential of the radion-inflaton system,
 \be
V(\sigma_v,\phi) =  e^{-\alpha\sigma_{v}/M_{p}}V(\phi) + U(\sigma_v)\,.
 \ee
Even though the global minimum do correspond to $\sigma_v=0$, 
it is easy to see  that during the phase of inflation driven 
by the inflaton potential the radion is trapped in a 
a false (effective) minimum. 
Indeed, if one takes the inflaton potential  as effectively constant, 
the minimization condition for the total radion potential reads
 \be 
  {\partial U(\sigma_v)\over \partial\sigma_v} = 
  {\alpha\over M_P} e^{-\alpha\sigma_{v}/M_{p}}V(\phi)\,.
 \label{falsemin}
 \ee
Clearly, $\sigma_v =0$ is not longer a solution, unless $V(\phi)$ were null. 
Thus, the minimum for the radion 
is displaced~\cite{kolb} to a position which  strongly
depends on the actual form of the radion potential. 
We should notice, however, that this effective minimum  
is always located on the positive range of $\sigma_v$ values, 
which means that within a Hubble time
the bulk has grown larger than its expected value (see
the definition of the bulk radion in Eq.~(\ref{radions})). 
This can be understood by noticing that the global minimum of 
$\sigma_v$ is at zero, and the radion is originally coming down the 
potential from negative values, therefore, 
only for $\sigma_v>0$ one could have a positive slope for $U(\sigma_v)$, 
as it is needed to match the RHS of Eq.~(\ref{falsemin}). 
Moreover, this equation indicates that the effective minimum is located at a
$\sigma_v$ value where the slope of the radion potential catches up with 
a now suppressed inflaton potential.

To clarify these points, let us take for instance the  potential
$U(\sigma_v) = {1\over 2}m^2 \sigma_v^2$, 
where the radion mass is assumed to be lighter than the inflaton potential 
scale, $M_I$. 
Here $M_I$ is taken to be the inflation scale as given by
the inflaton potential alone (i.e. we use $V(\phi)= M_P^2 M_I^2$). 
In Figure 1 we have plotted the total radion potential 
for a particular case where $\alpha M_I/m =10$. 
There we have also depicted each term on the potential. 
Now one can easily sees why and where the minimum gets displaced.
First of all the inflaton part of the potential drops very fast due to the 
exponential suppression of a growing radion, however, this is not fast
enough as to catch with the radion potential which, being quite flat, 
grows much more slowly. Next point is that the exponential has no 
minimum, so the minimum appears only due to $U(\sigma_v)$, and therefore, 
such a minimum should be close to the $\sigma_v$ value where  both the
potentials are almost equally relevant. Unfortunately, for the case we 
are considering, this happens when $\sigma_v$ is about three times 
larger than $M_p$. This effect is more dramatic when the hierarchy 
among $m$ and $H_I$ is larger. In such a case the inflaton potential 
in our picture is shifted to the right, which means that the minimum 
should be even at a larger $\sigma_v$ value.

\begin{figure}
\centerline{
\epsfxsize=250pt
\epsfbox{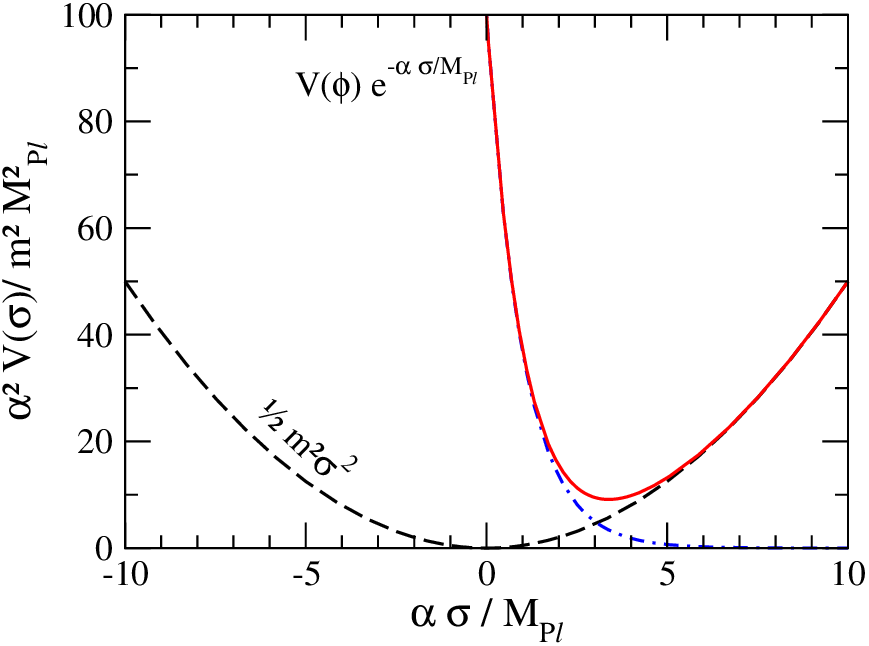}
}
\vskip1ex

\caption{Total radion potential for $\alpha M_I/m =10$ (continuous curve).
Dashed curve corresponds to the radion mass potential whereas 
the dash-dotted curve plots  the inflaton-radion coupling term. 
Notice the minimum of the total potential
occurs close to the point where both terms are of same order.}
\end{figure}

This is troublesome for our previous picture of inflation: 
it means that, although the inflaton is slow rolling, its contribution to 
the Hubble expansion is being affected by the fast evolution of the
exponential suppression that comes with a growing radion. 
Once both the terms in the potential are comparable we cannot neglect the
radion contribution to inflation anymore, which means that a third phase 
of inflation would take place. In this new phase both fields contribute 
to expansion, however, now Hubble scale would be much smaller than 
$M_I$ by some orders of magnitude. 
In fact, by using Eq.~(\ref{falsemin}), we find that  the equation for the 
minimum  satisfies 
 \[ m^2\sigma_v = {\alpha\over M_P} e^{-\alpha\sigma_{v}/M_{p}}V(\phi)\,.\]
Using this expression one can evaluate the whole potential at the effective
minimum and we obtain~\cite{kolb} 
 \be 
 V(\sigma_v) = m^2\sigma_v \left[{1\over 2}\sigma_v +{M_P\over\alpha}\right]\,.
 \label{Veff}
 \ee
In other words, the scale of inflation should  now be $H_{inf}\sim m$.

The system will remain in the false vacuum as long as the inflaton remains
constant.  Inflaton was, nevertheless, slow rolling right before the radion 
field settles  down to the false vacuum. As inflation proceeds, the
inflaton slowly takes smaller vevs, as does the energy density residing 
in its potential. For small values of the scale of the $V(\phi)$ potential, 
$M_I$, the minimum gets shifted to smaller values of $\sigma_v$. It would 
look like if the radion were slowly
rolling down an effective potential given by Eq.~(\ref{Veff}). By looking at 
the RHS of Eq.~(\ref{bulkexp}) we find that  the exponential suppression
of the inflaton potential does contribute for a very slow evolution of the
inflaton field, which behaves as if it were effectively frozen. 
All these would result in a large period of inflation, 
that may last plenty of e-foldings. 
As inflation continues in 4D,  the radion rolls down  and the
extra volume now  shrinks. It ends only  when  the inflaton 
leaves the slow rolling phase, by then it is very likely that the  
radion would still have a large, almost Planckian vev. Therefore, 
just after inflation ends the radion would again start oscillating with 
an amplitude of order of the Planck mass. Therefore bulk inflation may 
not help to completely solve the radion problem. 

\begin{figure}
\centerline{
\epsfxsize=250pt
\epsfbox{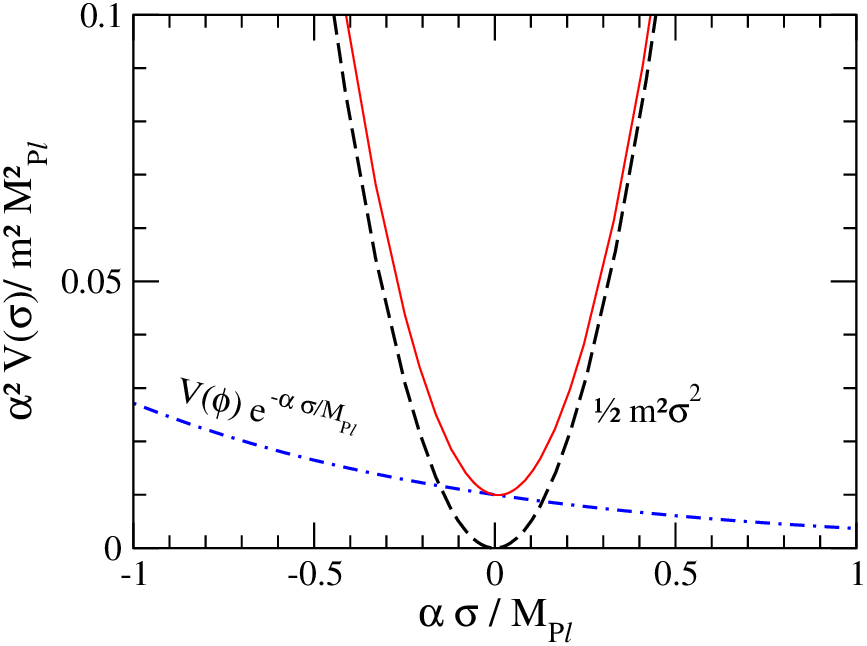}
}
\vskip1ex

\caption{ Same as before but now taking  $\alpha M_I/m =0.1$.
Notice that here the   minimum is consistent with $\sigma_v=0$.}
\end{figure}

On the other hand, this problem does not appear if
the radion mass were larger than the scale of the inflaton potential. 
Figure 2 plots the radion potential when $\alpha M_I/m = 0.1$.
Now the whole picture changes dramatically, the radion mass dominates on the
dynamics of expansion in a first phase. If the mass is large enough the radion 
quickly goes down the potential whereas the inflaton remains frozen. 
At some point, when the radion amplitude is quite small in the picture, the
inflaton takes over and drives the radion towards the effective minimum, which
is nicely just around $\sigma_v =0$. From there on the evolution would be
dominated by the inflaton potential and our discussion in the previous
subsection applies.

\subsection{A potential resolution to the radion problem}

Finally, let us consider what happens if the radion stabilization potential 
were similar to the one described in Eq.~(\ref{guess}):
 \be
 U(\sigma_v)\sim m^2_{v}M_{p}^2\left(1-e^{\alpha_v\sigma_{v}/M_{p}}\right)^2\,.
 \ee
We use this potential and plot the total radion potential in Figure 3 for the
same hierarchy of parameters used in Figure 1.
Now, the stabilization potential grows 
exponentially for positive $\sigma_v$ values and then catches up 
with the inflaton potential not far from $\sigma_v=0.5~M_P$.  
In contrast to what happened in the case depicted in
Figure 1, now the bulk volume is not allowed to grow too much beyond the 
stabilized value and this ameliorates the radion problem.
Notice also that at the effective minimum the total potential 
is still dominated by the inflaton, therefore, it gives the features 
we were taking as granted in the previous subsections. Once the radion 
is settled at the effective minimum expansion would be driven only by 
the inflaton. This kind of potentials gives an idea of what one 
would desire for the actual stabilization potential. Therefore we
conclude in a cheering note that indeed if the correct radion potential
is found we would be able to stabilize the radion to its true minimum 
during inflation, and thereby solve the radion problem.
\begin{figure}
\centerline{
\epsfxsize=250pt
\epsfbox{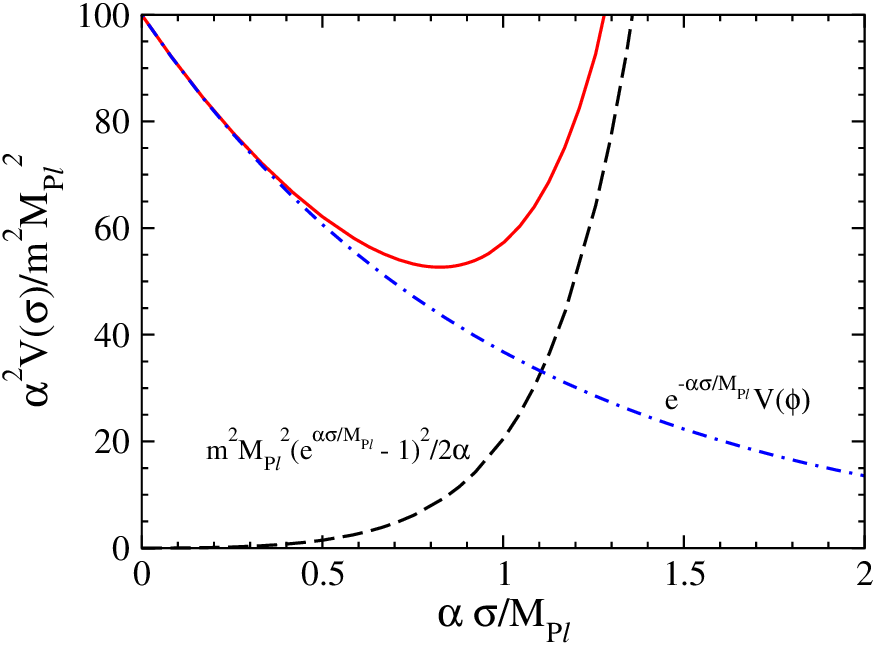}
}
\vskip1ex

\caption{The total radion potential as seen when the 
radion stabilization potential $U(\sigma_v)$ becomes rather steep 
for positive values of $\sigma_v$.
Here we have used  $\alpha M_I/m =10$. Notice that now the false minimum is 
closer to zero than before.}
\end{figure}

\subsection{Reheating}
   
It is believed that the total energy density of the inflaton is  transferred
into  radiation. The minimal requirement of cosmology is to have a thermal
bath with a
temperature more than ${\cal O}(1)$~MeV in order to preserve the successes  
of BBN. The final reheat temperature depends on the decay rates
$\Gamma_{\phi,N}$ of the oscillating field
\begin{equation}
T_{\rm r} \sim 0.1 \sqrt{(\Gamma_{\phi}+\Gamma_{N})M_{\rm p}}\,.
\end{equation} 
Note that in our model the bulk fields couple to effective $4$ 
dimensions via Planck suppressed interactions with the  
SM fields on the observable brane. 

However the zero mode inflaton can reheat the brane and also the bulk. 
This possibility was already addressed in Ref.~\cite{Mohapatra}
for the case of gravitons. From Eq.~(\ref{hybp0}) we found that 
both the fields obtain effective mass terms 
$\bar m_{\phi} \sim \bar m_{N} \propto \sqrt{2}\lambda N_0$, if we assume
$\lambda = g$. Setting $N_0 \sim {\cal O}(\rm TeV)$, then both $\phi$ and $N$
are kinematically allowed to decay into SM Higgs field, $h$. The decay rate  
can be estimated as follows~\cite{Mohapatra} 
\begin{equation}
\label{decay0}
\Gamma_{\phi, N \rightarrow hh}
	\sim \frac{f^2 M_{\ast}^4}{32\pi M_{\rm p}^2 m_{\phi,N}}\,,
\end{equation}
where $f$ is the coupling constant that we take of order one. 
If $\bar m_{\phi}\sim \bar m_{N} \approx 0.1M_{\ast}$, 
then the reheat temperature is $T_{\rm r}\leq 100$~MeV, 
for $M_*\sim$~100~TeV. 
Note that such a low reheat temperatures is a generic prediction if the
inflaton field is living in the bulk.

The next point is concerning the possible production of KK gravitons by the
inflaton. 
The KK gravitons can be directly produced from the decay of   
KK modes of 
$\phi$ and $N$, through the channel $\phi_{n} \rightarrow \phi_{l}G_{n-l}$,
where $n,l$ are the KK numbers, and $G$ is the KK graviton. 
On the other hand $\phi_{n}$ modes can be produced via collision processes, 
such as $\phi \phi \rightarrow \phi_{n} \phi_{-n}$, 
and  similar reaction for $N$.
The rate for exciting  KK inflaton modes goes as
 \begin{equation}
 \sigma_{\phi\phi \rightarrow \phi_{n}\phi_{-n}}
 \sim \lambda^2 \frac{M_{\ast}^2} {M_{\rm p}^4}\,,
 \end{equation}
whereas the   
total decay rate of a KK scalar decaying via graviton emission 
is estimated as
\begin{equation}
\Gamma_{\phi_n, total} = 
\sum_{\ell<n} \Gamma_{\phi_{n}\rightarrow \phi_\ell G_{n-\ell}}
\sim \frac{m_{n}^3}{12\pi M_{\ast}^2}\,,
\end{equation}
where $m_{n}^2=\bar m_{\phi}^2+n^2/R^2$. 
Note that the
$NN$ or $\phi\phi$ scattering rates for producing their KK counterparts are
more Planck suppressed 
than the direct decay of $\phi,~N$ to the brane fields, i.e. 
$\Gamma_{\phi\rightarrow hh} \gg 
\sigma_{\phi\phi \rightarrow  \phi_{n}\phi_{-n}}\times M_*^3$; 
where we have taken the number density of modes $\sim M_*^3$.
So, we find that the zero modes of $\phi$ and $N$ still prefer the Higgs as 
a final decay product. On the other hand,
the excited KK inflaton or KK partners of other scalar fields
present in the spectrum are extremely short lived. 
The heavier KK mode decays into the lighter KK modes plus gravitons, 
and eventually all the KK modes of the inflaton decay into the zero mode.
In case where extra bulk fields are present the reheating of the 
bulk can be naturally avoided, if 
either those modes are as heavy as the inflaton, or the effective 
inflaton couplings to the bulk fields are smaller than the 
inflaton-brane interactions. 

\section{Density perturbations}

\subsection{Converting radion isocurvature fluctuations to adiabatic perturbations via radion dominance}

We recognize that in a bulk inflation scenario the reheat temperature
of the universe is small, $T_{r}\sim 1-10$~MeV. The number of
e-foldings required for the COBE normalization depends on thermal
history of the universe after the end of inflation. For a low scale
inflation and low reheating temperature the require number of
e-foldings is small compared to the usual $60$ e-foldings, it is
roughly $43$ e-foldings~\cite{Lorenzana1,Green}.  As we discussed in
our earlier sections, the last $43$ e-foldings can easily arise during
the slow phase transition for natural parameters, e.g.  $N_{0}\sim
M_{\ast},~m_{\phi}\sim (10^{-1}-10^{-2})M_{\ast}$. However as it was
also pointed out in Ref.~\cite{Green}, the amplitude of the scalar
density perturbations during the phase transition evolves gradually to
become very small. Numerical studies suggest that for the above
mentioned parameters the amplitude of the perturbations is way too
small, e.g.  ${\cal P}^{1/2}_{H}< 10^{-15}$~\cite{Green}. The only way
this situation could be ameliorated if there is a large hierarchy
between the scales, e.g. $N_{0}\ll M_{\ast},~m_{\phi}\ll
M_{\ast}$. The summary is that we do not gain on density perturbations
by introducing two fields either. Nevertheless we should study other
interesting options, such as radion induced adiabatic perturbations.

Note that in our case $\sigma_{v}$ can be dynamically stabilized during
bulk inflation. However $\sigma_{\bot}$ cannot be dynamically driven to 
its minimum. As a consequence the actual dynamics of $\sigma_{\bot}$ can 
be treated independently. If $\sigma_{\bot}$ is sufficiently light compared
to the Hubble expansion during inflation then quantum fluctuations along
$\sigma_{\bot}$ direction generates isocurvature fluctuations, which could 
be transferred into adiabatic fluctuations when $\sigma_{\bot}$ decays while
dominating the universe. However we have to keep in mind that $\sigma_{\bot}$
must decay before BBN and it must dominate the energy density of the universe
while decaying. In case $\sigma_{\bot}$ dominates later on, then when it 
decays into SM degrees of freedom, it will inject significant entropy in 
the universe which will dilute the light nuclei created during BBN. 
Note that $\sigma_{\bot}$ decays only
via Planck suppressed interactions. This leaves us with a narrow window 
for the mass range of $\sigma_{\bot}$, which has to be large, e.g 
$m_{\sigma_{\bot}}=500$~TeV.

Important issue  to raise is what should be the amplitude of
isocurvature density perturbations produced by $\sigma_{\bot}$. Let us
for the time being concentrate on flat radion potential
$m^2_{\sigma_{\bot}}\sigma_{\bot}^2$. As we discussed in section
seven, see Eq.~(\ref{iso}), the amplitude of $\sigma_{\bot}$ during
that last $43$ e-foldings of inflation must be $10^{5}\times
M_{\ast}$.  If we set an opportunistic value,
$m_{\sigma_{\bot}}= 5\times10^{5}$~GeV and $M_{\ast}\sim 10^{6}$~GeV, then
$\sigma_{\bot,~inf}\sim 10^{11}$~GeV, in order to satisfy the
COBE normalization. Obviously, depending on the number of extra
spatial dimensions, if the initial vev of $\sigma_{\bot}$ at the onset
of inflation is $\sigma_{\bot}(0)\sim - 10~M_{p}$, see section
six, then following Eq.~(\ref{eqmot}) we can evaluate the approximate
number of e-foldings required to bring down its vev to a value when
the the interesting scalar perturbation modes are leaving the horizon.
The number of e-foldings comes out to be roughly ${\cal N}_{e}\sim
3\times10^{5}$. Note that here we are required to relax TeV as a
fundamental scale. The main bound comes from the radion mass
which has to decay before BBN in order to convert isocurvature 
perturbations to adiabatic ones. Also note that radion has to dominate
the energy density before decaying. This we will discuss below.

The conclusion is that if the fluctuations of $\sigma_{\bot}$
generates ${\cal P}^{1/2}_{\sigma_{\bot}}\sim 10^{-5}$, then there has
to be a large number of e-foldings of inflation. In this respect any
chaotic inflation model can do the job, it could either be a single
field inflation, e.g.  $V\sim m^2\phi^2$ with a large initial vev of
$\phi\sim 10^2-10^{3}\times M_{p}$, or two field hybrid inflation
model with a very slow phase transition, see Eq.~(\ref{hybp0}).
  
The spectral index of the fluctuations can be determined by
\begin{equation}
\label{tilt}
n-1\equiv \frac{d\ln{\cal P}_{\sigma_{\bot}}}{d\ln~k}=2\frac{\dot H_{\inf}}
{H_{\inf}^2}+\frac{2}{3}\frac{m^2_{\sigma_{\bot}}}{H^2_{\inf}}\,.
\end{equation}
If $\dot H_{\inf}\ll H^2_{\inf}$, which obviously depends on the bulk 
inflaton potential, then there is a slight dependence on the spectral 
index given by $n-1\sim (2/3)(m_{\bot}^2/H_{\inf}^2)$. On a generic 
ground one can say that if $m_{\sigma_{\bot}}< H_{\inf}\sim M_{\ast}$,
then the spectral index is very close to Harrison Zeldovich case,
e.g. $n\sim 1$. This is one of the signatures of radion induced 
fluctuations. Note that the energy density of the radion is always 
subdominant during inflation, e.g. $V> m^2_{\sigma}\sigma_{\bot}^2$.

However it is important that isocurvature perturbations stored in 
$\sigma_{\bot}$ must be converted into the adiabatic ones. This can
happen only if $m^2_{\sigma_{\bot}}\sigma^2_{\bot}$ dominates at later 
stages, but before BBN era. Therefore, the radion must be responsible 
for finally reheating the universe. However note that in our case 
this can happen soon after the end of inflation, because when 
$m_{\sigma_{\bot}}\sim H(t)$, the radion starts oscillating with an 
amplitude which is not far from $10^{8}$~GeV. Note that radion like 
the inflaton also couples to the SM degrees of freedom with a Planck 
suppressed interaction. As in our case both the inflaton and radion 
are oscillating simultaneously but with different frequencies, e.g.
$m_{\phi}>m_{\sigma_{\bot}}$. However inflaton being heavier will 
preferably decay before the radion and towards the last stage, if the 
radion oscillation could dominate the inflaton decay products, then, it 
is possible to convert isocurvature fluctuations into the adiabatic ones.
Otherwise there will be a contribution from isocurvature fluctuations also.
The total curvature perturbations is given by the sum of 
\begin{equation}
{\cal P}^{1/2}\sim f{\cal P}_{\sigma_{\bot}}^{1/2}+(1-f)
{\cal P}_{\gamma}^{1/2}\,,
\end{equation}
where $f$ is the fraction of energy density stored in the radion,
it is given by
\begin{equation}
f\equiv \frac{3\Omega_{\sigma_{\bot}}}{3\Omega_{\sigma_{\bot}}+
4\Omega_{\gamma}}\,.
\end{equation}
where $\Omega_{\gamma}$ is the energy density stored in the decay products of 
the inflaton. In our case this ratio could be small if 
$\Omega_{\gamma}>\Omega_{\sigma_{\bot}}$, especially since 
$V_{\inf}\gg m^2_{\sigma_{\bot}}\sigma^2_{\bot}$ and both the 
fields have a similar decay rate to the SM fields.
However if we imagine that not all
the inflaton energy density goes into the brane degrees of freedom, 
suppose the inflaton couples strongly to the bulk degrees of freedom, 
then it is quite possible that $\Omega_{\bot}\gg\Omega_{\gamma}$. Note 
that even though there is a larger relativistic brane degrees of freedom,
but the bulk volume is also large. Therefore it is possible to imagine
that a branching ratio between  brane and bulk is suppressed by the bulk 
volume factor.

The conclusion is that it is indeed possible to generate adiabatic 
density perturbations from the isocurvature fluctuations of $\sigma_{\bot}$.
However a realistic scenario requires larger fundamental scale than TeV,
e.g. $M_{\ast}\geq 500$~TeV.   

\subsection{Converting radion isocurvature fluctuations to 
adiabatic perturbations via fluctuating inflaton coupling}

There is an alternative mechanism for converting isocurvature
fluctuations of $\sigma_{\bot}$ to adiabatic perturbations from
fluctuating inflaton coupling. Note that in our case $\sigma_{v}$ has
a natural coupling to the inflaton while $\sigma_{\bot}$ is an
independent field which only couples SM degrees of freedom. Suppose
the $\sigma_{v}$ has not settled down to its
global minimum during inflation, which may occur if $\sigma_{v}$ has a
flatter potential for both positive and negative vevs. In such a case
the inflaton coupling to the SM fields will fluctuate when the
inflaton decays due to the presence of the term
$e^{-\alpha\sigma_{v}/M_{p}}V$ in the action. During inflation
$\sigma_{v}$ has fluctuations, and if it is lighter compared to
$H_{\inf}$ then it also has a non-zero vev. Further the above
term can give rise to fluctuating inflaton coupling or fluctuating
inflaton mass. In either case the perturbations can be transferred to
the SM degrees of freedom during the decay of the inflaton. This can
be understood in two steps; first note that fluctuating coupling or
mass gives rise to fluctuations in the decay rate of the inflaton and
second, fluctuating decay rate gives rise to fluctuations in the
reheat temperature of the universe. The fluctuations in $\Gamma$ can
be translated into fluctuations in the energy density of a thermal
bath with~\cite{Dvali} 
\begin{equation}
\frac{\delta\rho_{\gamma}}{\rho_{\gamma}}=-\frac{2}{3}\frac{\delta\Gamma}
{\Gamma}\,.
\end{equation} 
The factor $2/3$ appears due to red-shift of the modes during the decay 
of the inflaton whose energy still dominates. It has already been proven that
the curvature perturbation can be fed by the fluctuations in the decay rate
of the inflaton ${\cal P}^{1/2}_{\zeta}=-(1/6)\delta\Gamma/\Gamma$
\cite{Dvali,Marieke}. Note that in this case, the tilt in the spectral
index is given by Eq.~(\ref{tilt}).

In our case, fluctuations in the decay rate is given by
\begin{equation}
\frac{\delta \Gamma}{\Gamma}=-\frac{\delta\sigma_{v}}{\sigma_{v}}
\sim \frac{H_{\inf}}{2\pi\sigma_{v}}\,.
\end{equation}
In order to match the observed COBE normalization, the vev of 
$\sigma_{v}$ during inflation must satisfy 
$\sigma_{v}\sim 10^{5}H_{\inf}\approx 10^{5}M_{\ast}$. Since we know
the initial vev of $\sigma_{v}$ at the onset of inflation has to be 
around $\sigma_{v}(0)\sim 10~M_{p}$, therefore one requires
at least ${\cal N}_{e}\sim {\cal O}(10^{4}-10^{5})$ number of e-foldings 
to bring down the vev of $\sigma_{v}$ to appreciable value in order to
generate the right amplitude for the density perturbations.

In reality both the radions can play roles simultaneously; fluctuations in
$\sigma_{v}$ can give rise to fluctuations in reheat temperature of the
universe and late decay of $\sigma_{\bot}$ can transfer its isocurvature 
perturbations to adiabatic ones. However the dominant one will prevail and
it seems that both requires large number of e-foldings of inflation, 
which can be obtained for the bulk inflation only, see Eq.~(\ref{hybp0}).

\section{Radion as a candidate for Cold Dark matter}

Since the class of models we are discussing do not have the conventional
dark matter candidates such as the axion or the neutralino, we like to
explore other particles as CDM. As we have noticed
a successful cosmological
scenario seems to point to a reheat temperature of the universe which is
much smaller than the cut-off scale $M_{\ast}$. Therefore in order
to have a stable KK dark matter\cite{tait}, it is necessary to excite them 
non-thermally. Nevertheless we should also keep in mind
that we do not understand thermalization of the inflaton decay very well
and it could be that for some reason the reheat temperature is close to
$T_{r}\leq M_{\ast}$ in which case KK modes would have no trouble getting
excited and play the role of dark matter.

There however exists another interesting possibility that the radion
oscillations 
can also act as a cold dark matter provided the radions are absolutely 
stable and light. According to Eq.~(\ref{taur})
this happens if the radion mass is lighter than 
30~MeV. Thus, if we believe in a radion potential that traps 
the radion in the minimum during inflation, then once 
$m_{\sigma_{v}}\sim H(t)\ls 30MeV$ or 
$m_{\sigma_{\bot}}\sim H(t)\ls 30 MeV$, the 
radions start oscillating respectively in their minimum with an 
amplitude of order $H(t)$. These are coherent 
oscillations which give rise to an effective pressureless fluid and
the density stored in their coherent oscillations can give 
rise to a cold dark matter. This issue will be dealt separately
in near future.

\section{Conclusion}

In this paper we have studied some aspects of the cosmology of  
models with universal extra dimensions, where SM particles live 
on a $4+n$ dimensional brane, whereas gravity resides in $4+n+p$
dimensions (with $p$ dimensions being much larger than $n$ dimensions).
These models generically have two radion fields in the effective
four dimensional theory.
The two radions can be identified as
(i) The volume radion, associated to the variation of the whole 
volume of the extra space; and (ii) the shape mode, 
produced by the variations in the hierarchy among the $n$ small 
and the $p$ large extra dimensions in the theory. 

Our study shows the difficulties that the radions introduce in the 
description of early universe cosmology. In exploring this we consider
two extreme forms of the radion potential. If the radion stabilization
potential is 
rather flat, almost all scenarios of inflation suffer from a  
radion problem analogous to the well known moduli problem. We note that, 
although the radions themselves can drive inflation, the picture 
seems hard pressed to reconcile with the correct amplitude for the
the primordial density perturbations and, typically inflation ends 
leaving the radions to oscillate with large amplitudes. We then focus on
brane induced inflation and note that
 inflation occurs at relatively low scale and the inflaton 
energy density is typically too small (about ${\cal O}({\rm eV})^4$) 
to drive a successful inflation. On the other hand, a bulk inflaton can
help to settle down the volume radion 
to an effective minimum within a Hubble time and seems more promising
from the point of view of density perturbations. 
Nevertheless, the effective  minimum for the radion potential is
usually located at large values of the radion field, which means that the 
bulk inflaton makes the volume of the extra dimensions to grow much 
beyond the expected stabilized size. If inflation is supported by 
both the volume radion and the inflaton, while the inflaton keeps 
rolling down the volume shrinks. In contrast,  the
shape mode gets decoupled and therefore remains frozen during the entire
inflationary phase. By the end of inflation it is very likely that both 
the radion modes will end with large amplitudes, unless
the universe  passes through an extremely large number of e-foldings.

Most of the above mentioned problems get resolved  if the 
radion mass is assumed to be large. 
This allows the radion to decay fast into SM particles while oscillating
at late stages. With a bulk inflaton, the radion problem can  
disappear completely for the volume radion which is
now rapidly driven to its actual minimum and trapped there by the dynamics of
inflation. For this it is mandatory that the flatness of the volume
modulus is lifted by some brane or bulk physics. We have explored
phenomenologically this possibility.

Finally we concentrated on the radion induced isocurvature fluctuations.
The observations demand that we convert this isocurvature fluctuations
to the adiabatic ones. It is quite possible in our case because the
radions can decay into the SM degrees of freedom. However this again 
would require
the radion masses to be fairly large $\geq 10$~TeV at least, so they decay
before BBN. Another interesting possibility which we have noticed is that 
the inflaton coupling to the SM matter can fluctuate, because of the couplings
being governed by the vevs of the radions. These fluctuations induce 
fluctuation in the coupling and therefore in the reheat temperature of 
the Universe. This could be an interesting avenue to obtain the desired
amplitude for the density perturbations. This particular scenario does not
demand large masses for the radions, but it certainly requires fairly 
intermediate scale vevs for the radion mass, e.g. $\sim 10^{5}M_{\ast}$.
This can be obtained easily if there is large e-foldings of inflation,
which occurs very naturally in the bulk driven inflation scenario.

The punch line of this paper can be summarized as follows; 
for ${\cal O}(100)$~TeV scale gravity the bulk driven inflation 
along with lifted potential for the volume radion can give rise to 
a successful cosmology.

\acknowledgments

A.P.L. would like to thank the Particle Theory group of the University
of Maryland for the warm hospitality during the first stages of this
work. The works of R. N. M. and partly of A.P.L. are supported by the
National Science Foundation Grant No. PHY-0099544. A. M. is a CITA
national fellow.



\end{document}